\documentclass[prd,aps,nofootinbib,superscriptaddress,twocolumn]{revtex4}

\usepackage{graphics}
\usepackage{amsmath}
\usepackage{epsfig}
\usepackage{wasysym}

\begin{document}

\title{Dark Matter in Minimal Universal Extra Dimensions\\ with a Stable Vacuum and the ``Right'' Higgs}

\author{Jonathan M. Cornell}
\email{jcornell@ucsc.edu}\affiliation{Department of Physics, University of California, 1156 High St., Santa Cruz, CA 95064, USA}\affiliation{Santa Cruz Institute for Particle Physics, Santa Cruz, CA 95064, USA} \affiliation{Oskar Klein Centre for Cosmoparticle Physics, Department of Physics, Stockholm University, SE-106 91 Stockholm, Sweden}

\author{Stefano Profumo}
\email{profumo@ucsc.edu}\affiliation{Department of Physics, University of California, 1156 High St., Santa Cruz, CA 95064, USA}\affiliation{Santa Cruz Institute for Particle Physics, Santa Cruz, CA 95064, USA} 

\author{William Shepherd}
\email{wshepher@ucsc.edu}\affiliation{Department of Physics, University of California, 1156 High St., Santa Cruz, CA 95064, USA}\affiliation{Santa Cruz Institute for Particle Physics, Santa Cruz, CA 95064, USA}

\date{\today}

\begin{abstract}
The recent discovery of a Higgs boson with mass of about 125 GeV, along with its striking similarity to the prediction from the Standard Model, informs and constrains many models of new physics. The Higgs mass exhausts one out of three input parameters of the minimal, five-dimensional version of universal extra dimension models, the other two parameters being the Kaluza-Klein (KK) scale and the cut-off scale of the theory. The presence of KK fermions with large coupling to the Higgs implies a short-lived electro-weak vacuum, unless the cut-off scale is at most a few times higher than the KK mass scale, providing an additional tight constraint to the theory parameter space. Here, we focus on the lightest KK particle as a dark matter candidate, and investigate the regions of parameter space where such particle has a thermal relic density in accord with the cosmological dark matter density. We find the paradoxical result that, for low enough cutoff scales consistent with vacuum stability, larger than previously thought KK mass scales become preferred to explain the dark matter abundance in the universe. We explain this phenomenon by pinpointing the additional particles which, at such low cutoffs, become close enough in mass to the dark matter candidate to coannihilate with it. We make predictions for both collider and direct dark matter searches that might soon close in on all viable theory parameter space.
\end{abstract}

\maketitle

\section{Introduction}

The dark matter in the universe is the strongest piece of evidence for new particle physics beyond the Standard Model. Dark matter has been inferred from the kinematic properties of structures as diverse in scale as galaxy clusters down to small dwarf galaxies within the Milky Way halo and direct evidence for dark matter has been deduced in gravitational lensing observations. The effects of dark matter on early universe cosmology have also been measured to exquisite detail, most recently by the Planck collaboration, giving, in conjunction with other cosmological data, a very precise measurement of the energy density of dark matter in the universe.

The single most compelling explanation of the dark matter from a particle physics standpoint is a category of particle models known as ``weakly interacting massive particles'', or WIMPs \cite{wimp_miracle}. If the universe proceeded thermally from some high temperature to today without any major disruptions (such as episodes of entropy injection), the thermal relic density expected of a particle with a mass-scale near the electroweak scale, and with electroweak interactions, would be close, at least in order of magnitude, to the observed density of cosmological dark matter. Interestingly, and, to some investigators, ``miraculously'', many models designed to address entirely different particle physics issues accidentally include such a ``WIMP-y'' particle, which makes this class of particle dark matter models even more compelling.

One key theoretical question at the interface of cosmology and particle physics is why the universe is composed of four space-time dimensions. While many different approaches have been taken to this question, a particularly interesting idea is to dispute the assertion itself \cite{Antoniadis:1990ew}. It is entirely possible that the universe possesses {\em extra} dimensions which, for whatever reason, cannot be readily detected with experiments. One way in which this could be achieved is to have extra dimensions into which Standard Model particles, for well-defined physical reasons, cannot propagate \cite{ArkaniHamed:1998rs}. Then we would hope to be able to measure the existence of the additional dimensions in gravitational experiments, and we might be able to detect new particles that do travel in the extra dimensions occasionally interacting with ordinary matter.

Even models where the entire Standard Model is constrained to not travel in the extra dimension are still subject to strong constraints from gravitational measurements, which implies that the extra dimensions have to be small \cite{PDG}. Measurements of the dependence of the force of gravity on distance have constrained any extra dimensions to be smaller than a fraction of a millimeter in size, and additional constraints from particle colliders producing gravitons traveling in the extra dimensions place even more stringent bounds \cite{ADDcollider}. Given that the extra dimensions must be so small, it no longer becomes imperative that SM fields be forced to not propagate in the bulk. Models where all Standard Model fields travel in the bulk are known as Universal Extra Dimensions (UED). From a four-dimensional perspective, fields propagating in the additional extra dimensions manifest themselves as Kaluza-Klein (KK) excitations, with a mass, to zero-th order, corresponding to the inverse-size of the extra dimension. Coincidentally, the lightest KK excitation is stable (due to a residual KK ``parity'') and could be a WIMP dark matter candidate \cite{uedreview}.

Any model of new physics beyond the standard model is now confronted by a new constraint stemming from the recently discovered Higgs boson \cite{Higgsdiscovery}. Many models have been designed to explain why the Higgs can be light in spite of quantum corrections which na\"ively are expected to push the particle's mass up to high scales; those models generically make predictions for what the Higgs mass would be if the corresponding mechanism is correct. Even models which do not address this hierarchy problem are at least required to match the experimental data pouring in regarding this new particle. For relatively simple models like minimal UED, the decrease in the viable parameter space leads to new predictive power: the parameter space of minimal UED is in fact reduced to being effectively two-dimensional.

The discovery of the Higgs and the measurement of its mass has additional implications for mUED models. In the Standard Model it appears that the electroweak vacuum is only metastable according to the central values of the measured Higgs and top quark masses, but it is actually still possible that the vacuum remains stable up to the Planck scale \cite{Degrassi:2012ry}. This behavior is, of course, modified in the presence of any new physics scenario intervening at some mass scale intermediate between the electroweak and Planck scales. In the context of the UED picture, the vacuum stability is  under significant additional stress by the presence of numerous KK fermions strongly coupled to the Higgs, leading to very significant bounds on the parameter space of the model \cite{vacuumstability}.

In this work we focus on the region of parameter space of mUED which is compatible with the Higgs mass and properties, with the requirement of a sufficiently long-lived electroweak vacuum, with direct collider searches for KK particles at the LHC, with the requirement of the lightest KK particle to have the correct thermal relic density and, finally, with such particle be compatible with direct dark matter searches.

In some of the previous studies, the  parameter space considered in light of UED dark matter phenomenology had been the KK scale $R^{-1}$ and the Higgs mass $M_h$, or $R^{-1}$ and the cut-off scale $\Lambda R$ for putative values of $M_h$ (see e.g. \cite{uedreview} and references therein). Now that the Higgs mass has been experimentally measured, we can consider the complete parameter space in one plot, and at the same time extract interesting implications from the vacuum stability requirement, since the ultra-violet (UV) cutoff of the theory is lowered, which is precisely where vacuum stability concerns force the theory to live. Shortly before this work was released, a compendium of recent UED constraints appeared \cite{Servant}. Our work contrasts with this one in our focus on the low values of the cutoff scale favored by the vacuum stability constraint, but our conclusions are qualitatively similar for the parameter space where our studies overlap.

This study is organized as follows. In section \ref{sec:mUED} we introduce and discuss the relevant features of the mUED model. We then discuss current bounds from collider searches in section \ref{sec:LHC}. We consider the dark matter physics of the model in section \ref{sec:DM}. Finally, we present our conclusions in section \ref{sec:conc}.

\section{Minimal UED}
\label{sec:mUED}

In Universal Extra Dimensions all Standard Model fields can propagate in a fifth (and possibly more) extra dimension(s). The extra dimension, of course, must be compactified to allow the model to reduce to the SM in the limit of low energy. In its most minimal incarnation, the extra dimension is compactified on a circle of radius $R$, which is then orbifolded to regain the possibility of having chiral fermions, leaving a line with specific adequate boundary conditions on the fields.

As in any model of small extra dimensions, the additional direction manifests itself in four dimensions as a tower of Kaluza-Klein excitations of the fields allowed to propagate in the bulk. The characteristic mass splitting between the KK levels is set by the inverse compactification scale $R^{-1}$. As higher-dimensional theories are not renormalizable, it is necessary that additional new physics be invoked at some scale to restore good behavior at high energies; generically this cutoff scale is specified in the natural units of the extra dimension as the unit-less number $\Lambda R$.

In principle, there remain many free parameters having to do with the exact nature of the boundary conditions at the orbifold points in the extra dimensions, and these cannot be consistently set to zero as they are induced at loop level, but the framework of minimal UED \cite{Cheng:2002ab} assumes that they are small enough to not significantly distort the spectrum or couplings of the theory, and neglects these parameters {\em tout-court}. With this assumption, the only remaining parameters needed to fully specify the model are the SM parameters, including of course the Higgs boson mass. The recent measurement of the Higgs mass \cite{Higgsdiscovery} thus removes the last unconstrained parameter in the theory parameter space, leaving us with a treatable two-dimensional parameter space to consider.

At tree level, the KK excited particles are all degenerate up to corrections from Standard Model particle masses, which with few exceptions are small in comparison to the KK scale $R^{-1}$. However, loop corrections to the masses can be important, and have been calculated in detail \cite{Cheng:2002iz}. The splittings grow logarithmically with the cutoff scale $\Lambda R$, as expected for a loop effect, with colored particles rising in mass more quickly than others, while the dark matter candidate (for the measured Higgs mass the KK level-1 excitation of the hyper charge gauge boson $B_1^\mu$) drops slightly below the KK mass scale $R^{-1}$. These splittings have very important consequences for expected signals at colliders and for dark matter physics.

The measured value for the Higgs mass presents an interesting conundrum, as it implies, at face value, that, at some very high energy, the potential for the Higgs boson develops a second minimum at very large field values, implying a second, lower-energy ``true'' vacuum that our universe could, in principle, decay into. The precise energy scale at which this takes place (and in fact the existence or absence of this vacuum state) depends very sensitively on the top quark mass as well as on the Higgs mass, but for all experimentally reasonable values the current universe is at least metastable. A viable model thus requires the lifetime of the universe's decay to be at least longer than the measured age of the universe itself.

In models with universal extra dimensions this problem becomes quite acute. The decrease in the Higgs quartic coupling which ultimately leads to this issue is driven largely by the top quark Yukawa coupling, and in UED models there are additional fermions with strong coupling to the SM Higgs, namely the KK excitations of the top quark. As a result, the quartic coupling runs much faster as one introduces more ``effective top quarks'', and vacuum stability becomes a strong bound on the model parameters. Constraints of this nature can always be taken as defining a scale at which new physics must somehow stabilize the observed vacuum, so these constraints can be directly translated into an upper limit on $\Lambda R$ for a given choice of $R^{-1}$. These constraints have been explored in detail \cite{vacuumstability}, and allow maximal values of $\Lambda R$ of about 5, relatively irrespective of the KK mass scale $R^{-1}$. We show and summarize these constraints on the relevant theory parameter space in Fig. \ref{fig:limits} with a dashed brown line. Values of $\Lambda R$ above the line imply a metastable universe with an unacceptably short lifetime.

\section{UED at the LHC}
\label{sec:LHC}

Theories of universal extra dimensions have sometimes been dubbed ``bosonic supersymmetry'' because of the striking similarities with the phenomenology of supersymmetric models, especially at colliders. It is not surprising, therefore, that searches which are currently placing some of the most stringent constraints on mUED parameter space are typically ``re-purposed'' supersymmetry (SUSY) searches. 

The first KK excitation level contains partners to all SM particles, much like a spectrum of superpartners. Generically, however, mUED predicts a much more compressed particle spectrum  than what one would expect from MSSM superpartners, particularly in the region of low $\Lambda R$ suggested by the properties of the Higgs and by the constraint of vacuum stability. As a result, cascade decays in mUED are much more experimentally challenging than in supersymmetry: all of the SM particles produced are much softer. Nevertheless, interesting bounds can be placed on the mUED parameter space in the simplest searches for squark pairs. 

Most of the bounds alluded to above come from initial state radiation (ISR) jets produced in association with the KK excitations. The ISR jets pass the cuts designed to catch the jets from the squark decay to quark and neutralino, and the momentum of the KK excitation is transferred almost entirely into the DM candidate. Current bounds from these searches are detailed in Ref.~\cite{jetlimits}, and are presented along side our main results in Fig. \ref{fig:limits}. They constrain the UED mass scale $R^{-1}$ to be greater than approximately 825 GeV, and are largely independent of $\Lambda R$.

Searches for SUSY which require leptons in the final state can, in principle, have more sensitivity to the cascade decays in mUED than those which use only jets, as soft leptons are much more rare in background than soft jets. The use of such searches to set limits on 6-D UED scenarios was discussed in \cite{6d}, while the ATLAS experiment itself has recast some of its SUSY searches in this light for mUED \cite{atlassusy}. These bounds, also shown in Fig. \ref{fig:limits}, exhibit an interesting behavior with $\Lambda R$. They are very weak for the extremely degenerate spectra at very small $\Lambda R\sim5$, and become most strong at $\Lambda R\sim 10$, where they bound $R^{-1}$ to be greater than about 800 GeV. They then weaken as the colored states become heavier with further increases in $\Lambda R$. These bounds are theoretically a bit more robust than those depending on purely hadronic final states as they are less sensitive to possible mis-modelling of the ISR jets crucial to the hadronic signatures. A search which has been specially designed to find mUED may have more reach to the lower values of $\Lambda R$, but the backgrounds at low particle momenta will remain very difficult to overcome.

There has been recent interest in attempting to directly bound the second KK excitations of particles \cite{Edelhauser:2013lia, Matsumoto:2009tb, Chang:2012wp}, as such particles can have loop-level coupling directly to SM fields, and thus may appear as resonances, thus much cleaner to detect than the soft decay products of the first KK level particles. These searches, however, suffer from the smallness of the loop-induced couplings. As a result, the resulting constraints are generically weaker than those directly searching for the first level KK excitations and we do not include them in our plots. We also note that recent works have shown that current LHC constraints on the loop driven effective coupling of the Higgs to gluons \cite{Dey:2013cqa} and to W bosons \cite{Datta:2013xwa} seem to limit $R^{-1} > 1.3$ TeV, but constraints on $R^{-1}$ from the limits on Higgs couplings to other particle species are far less stringent.

\section{Dark Matter in mUED}
\label{sec:DM}

The field orbifold boundary conditions break down momentum conservation in the extra dimensions to a discrete $Z_2$ symmetry known as KK parity: odd KK modes have KK parity opposite to SM particles, and as a result the lightest KK particle (LKP; here, $B^{\mu}_1$) is stable. This is similar to R-parity in supersymmetric models, albeit KK parity stems from a completely different physical origin. 

The relic density of the LKP has been calculated in several previous works, first by \cite{Servant:2002aq}. The key issue in this type of calculation is the presence of numerous particles freezing out at similar epochs to the LKP, given the mass-degenerate nature of the mUED spectrum. As a result, the inclusion of a more and more complete set of coannihilating partners \cite{threeexc} has been the name of the game in achieving the highest possible accuracy. 

The original calculation of Ref.~\cite{Servant:2002aq} was extended and refined in Ref.~\cite{Kong:2005hn, Burnell:2005hm}, who considered coannihilation processes with all first level KK partners and included all possible tree level (co-)annihilations into 2 SM particles. The next step in complexity arises from the fact that, again due to the nature of the KK ladder, KK level 2 states have a mass comparable to twice the mass of KK level 1 particles, including the LKP. Resonant annihilation (albeit suppressed by small KK level 1-1-2 couplings) is thus a potentially very important effect. This point was addressed  in Ref.~\cite{Kakizaki:2005en, Kakizaki:2005uy, Kakizaki:2006dz}, where it was pointed out that loop induced couplings of second level KK particles to a pair of SM particles are important for the relic density calculation, as such couplings lead to the mentioned resonantly enhanced annihilation processes. 

Finally, it was more recently pointed out, in Ref.~\cite{Belanger:2010yx}, that annihilation into second level KK states should be considered in the calculation of the relic density as well, as many of these states will subsequently decay nearly entirely to SM states via a loop process, effectively contributing to the total cross section into SM particles relevant to the freeze-out process. 

\begin{figure}%
\includegraphics[width=\columnwidth]{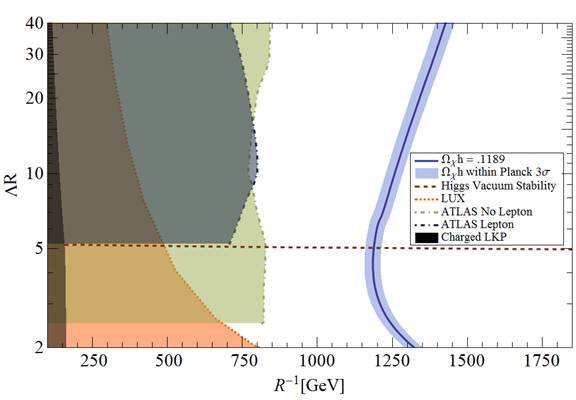}%
\caption{Current collider, direct detection, Higgs vacuum stability, and cosmological limits on the mUED parameter space.}%
\label{fig:limits}%
\end{figure}

In this work we include {\em all the mentioned layers of complexity} in calculating the LKP relic density, and use the ``right'' Higgs mass \cite{Higgsdiscovery}. Specifically, we use the most recent mUED CalcHEP model file discussed in \cite{Belyaev:2012ai}, modifying it to include all-loop level couplings of second level KK states to SM particles discussed in \cite{Belanger:2010yx}, and use it in connection with version 3.2 of the micrOMEGAs code \cite{Belanger:2006is,Belanger:2013oya} to calculate the LKP relic density. Our results are shown on the mUED parameter space ($R^{-1},\Lambda R$) by the shaded blue band in Fig.~\ref{fig:limits}.

As previously discussed, our study focuses on the ($R^{-1},\Lambda R$)  parameter space, as the Higgs mass is now known to high precision. In the calculation whose results are shown in  Fig. \ref{fig:limits}, we have assumed that the second level KK photon and the second level KK Higgs particles and photons decay completely to SM particles. This is driven mainly by the fact that the masses of the second KK excitations are near or below threshold for the decay into the appropriate two first KK excitations, and other decays are loop-suppressed to the same degree as those directly to a SM pair. Including these particles in the final state collects the most important contributions to 3-body SM final states LKP annihilation. 

In Fig.~\ref{fig:limits} the solid blue line corresponds to the best fit value for the dark matter relic density, $\Omega_\chi h = 0.1189$ \cite{Ade:2013zuv}, quoted by the Planck collaboration when including their cosmic microwave background (CMB) results, WMAP polarization results, high-$l$ CMB data from ground telescopes, and baryon acoustic oscillation measurements. The shaded region around this line represents the 3 $\sigma$ uncertainty on this result, corresponding to a range of values for $\Omega_\chi h$ between 0.1136 and 0.1238. All values of $R^{-1}$ greater than the value traced by this line are forbidden as they lead to over-closure of the universe.

We also include the recent collider limits discussed in the previous section, as well as constraints on the parameter space from direct detection experiments \cite{Aprile:2011hi}. For the latter, we have utilized spin independent cross sections as calculated with micrOMEGAs \cite{Belanger:2008sj} and the most recent exclusion limits at 90\% confidence level on these cross sections from the Xenon100 \cite{Aprile:2012nq} and LUX \cite{Akerib:2013tjd} collaborations.

Finally, as mentioned above we have included the limit found by \cite{vacuumstability}, requiring that the universe have a sufficiently stable electroweak vacuum. This is plotted as a dashed brown line in Fig. \ref{fig:limits}, and is approximately a constant upper bound on $\Lambda R < 5$ for all the mass scales of interest for dark matter physics.

One of the key findings of the present investigation is that the cosmologically favored value for the KK scale $R^{-1}$ {\em increases} at low $\Lambda R$ (the region of parameter space favored by vacuum stability constraints) as $\Lambda R$ decreases, while the behavior is the opposite for higher values of $\Lambda R$. Inspecting the relevant processes contributing to the LKP freeze-out, we find that this is to be attributed to coannihilation processes between level 1 KK photons and other level 1 KK particles at low values of $\Lambda R$, particularly the KK excitations of the gluon and quarks, which quickly become irrelevant as $\Lambda R$ increases because the splitting between the mass of the level one KK photons and colored KK particles grows rapidly.

\begin{figure}%
\includegraphics[width=\columnwidth]{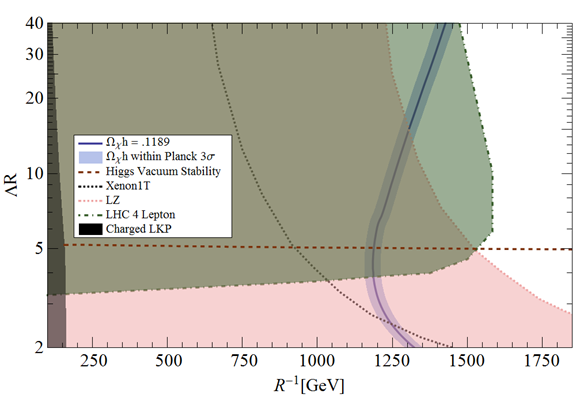}%
\caption{Projected collider and direct detection sensitivities in the mUED parameter space}%
\label{fig:projections}%
\end{figure}

In Fig. \ref{fig:projections}, we show the predicted sensitivities of the LHC and of future dark matter direct detection experiments in the mUED parameter space. The projected sensitivity of Xenon 1T is taken from \cite{Cushman:2013zza}, and the future LHC reach is based on a four-lepton search originally presented in \cite{Cheng:2002ab}, which has been updated to account for different possible splittings in \cite{Arrenberg:2013paa}. We note that, since this final state requires splittings large enough to detect the leptons produced in cascade decays of KK particles, sensitivity is lost as the smallest values of $\Lambda R$. These future bounds together leave only a small window of viable dark matter parameter space around $R^{-1}$ of 1225 GeV and $\Lambda R$ of 3. It is possible that future ISR-driven searches may have sensitivity to this region, but no explorations of that sensitivity which include the important systematic errors exist. Future generation-2 experiments such as DarkSide G2 and LZ will conclusively probe the entire range of parameter space, extending the reach of direct dark matter detection well beyond the cosmologically favored blue band \cite{Cushman:2013zza}, as shown by the pink dotted line.

We have also explored the effects on the calculation of the relic density of including the annihilation of the LKP to second-level KK states as well as the effect of including loop-induced vertices of second-level KK particles to SM particles. For a benchmark value of $\Lambda R = 20$, the cosmologically favored valued of $R^{-1} =$ 1340.8 GeV when the calculation is done as described in the previous paragraphs. However, the favored value of the compactification scale drops to 1105.0 GeV  when we do not take into account the decays of second level KK particles in the final state to SM states in our relic density calculation, and it drops further to $R^{-1} =$  1011.9 GeV when we then remove loop induced couplings of second level KK particles to SM particles from our CalcHEP model file. The effects of including these annihilation processes in the relic density calculation become more pronounced at increasingly higher values of $\Lambda R$. Therefore, our results are in agreement with those of Ref. \cite{Belanger:2010yx}: the inclusion of loop-induced couplings and particularly annihilation processes with second level KK particles in the final state make a substantial difference in the calculation of the relic density pushing the KK scale to up to more than 30\% higher, with obvious important phenomenological implications.

\section{Discussion and Conclusions}
\label{sec:conc}

We explored the parameter space of the minimal Universal Extra Dimensions framework after the Higgs discovery. We outlined and gave an updated overview on the collider and direct dark matter constraints on the relevant parameter space defined, now that the Higgs mass is known, by the inverse compactification scale $R^{-1}$ and the effective theory cutoff scale $\Lambda R$. These constraints effectively limit the KK scale $R^{-1}$ to values in excess of 700-800 GeV, depending on the precise value of $\Lambda R$, with direct detection searches covering most effectively the low $\Lambda R$ region where collider searches are less effective due to a highly degenerate mass spectrum.

The requirement that the electroweak vacuum be stable bounds the theory parameter space from above, setting an approximate upper limit to $\Lambda R < 5$ for mass scales that avoid overclosing the universe. We carried out the most accurate to-date calculation of the LKP relic density, and we found that the physics driving the LKP dark matter relic abundance in the low $\Lambda R$ region is richer than previously thought. Several coannihilation partners for the LKP dark matter particle contribute significantly to the total effective pair-annihilation cross section, and important effects arise also from resonant KK-level 2 modes as well as from pair-annihilation into KK-level 2 particles subsequently decaying into SM particles. The overall result is a significant {\em increase} in the cosmologically favored $R^{-1}$ range towards larger KK masses, as well as a non-trivial behavior with the cutoff scale $\Lambda R$. For intermediate values of $\Lambda R$, the benchmark range for $R^{-1}$ is around 1.2 TeV. We note that our results for the relic density constraints differ from \cite{Arrenberg:2013paa}, because of our inclusion of resonances and annihilations to second level KK states in the calculation, increasing by as much as 30\% the cosmologically favored value of $R^{-1}$.

Finally, we discussed prospects for the detection of a signal from mUED with both direct detection and collider experiments. We showed that the cosmologically favored parameter space will be entirely exhausted by generation-2 direct dark matter noble gas experiments such as DarkSide G2 and LZ, and LHC searches should also cover much, perhaps all, of the viable parameter space for dark matter in mUED.

\section*{Acknowledgments}
\noindent  JMC is supported by the NSF Graduate Research Fellowship under Grant No. (DGE-0809125) and by the Swedish Research Council (VR) and the NSF through the NSF Nordic Research Opportunity. WS and SP are partly supported by the US Department of Energy under contract DE-FG02-04ER41268.

\end{document}